\documentclass{article}

\usepackage{spconf,amsmath,amssymb,graphicx}
\usepackage{booktabs,array}
\usepackage{cite}
\usepackage{microtype}
\usepackage{tikz}
\usepackage{hyperref}
\usetikzlibrary{arrows.meta,positioning}

\hypersetup{
  hidelinks,
  pdftitle={When Is Inaction a Mistake? Continuation-Aware Auditing of PPO Trading Policies},
  pdfauthor={Xingfei Zeng, Xin Zhong, Nanting Li, Ziyang Zhong, Lei Xiao, Guanghui Lu},
  pdfkeywords={financial signal processing, reinforcement learning, Kalman filtering, transaction costs, policy diagnosis}
}

\graphicspath{{figures/}}
\newcommand{\NT}{\mathrm{NT}}
\newcommand{\WFI}{\mathrm{WFI}}
\newcommand{\OAR}{\mathrm{OAR}}
\newcommand{\AMA}{\mathrm{AMA}}
\title{When Is Inaction a Mistake?\\
Continuation-Aware Auditing of PPO Trading Policies}

\name{Xingfei Zeng$^{1,\dagger}$,
Xin Zhong$^{2}$,
Nanting Li$^{1}$,
Ziyang Zhong$^{1}$,
Lei Xiao$^{1}$,
Guanghui Lu$^{1,*}$}
\address{$^{1}$University of Electronic Science and
Technology of China\\
$^{2}$The University of Tokyo\\
\texttt{xingfeizeng@gmail.com; ghlu@uestc.edu.cn}\\
$^\dagger$First author;\quad
$^*$Corresponding author}

\begin{document}
\ninept
\maketitle

\begin{abstract}
An optimal reference may recommend trading when a learned policy chooses
inaction, but the recommendation depends on information and future decisions.
We introduce a four-stage audit for frozen proximal policy optimization
policies without retraining. It examines deployment occupancy, matches current information,
tests isolated deviations under incumbent continuation, and evaluates repeated
deployment of observation-based alternatives. In controlled linear--Gaussian
simulations, information matching explains part of the disagreement, while
continuation changes its interpretation. At unit observation noise,
incumbent continuation reverses 99.3\% of projected-hard missed-advantage mass;
repeated projected-rule deployment improves all 50 policies. These comparisons
distinguish isolated action changes from policy replacement. Historical
Bitcoin/Tether (BTCUSDT) replay applies this deployment perspective to a
hand-specified intervention selected using 2024 data and frozen for 2025.
Daily net reward improves by 135.03 basis points, with gains in 46 of 50
policies, primarily through lower turnover costs. The audit clarifies
what oracle-flagged inaction implies for deployed decision making.
\end{abstract}


\section{Introduction}
Choosing not to act is essential when actions incur costs. In trading,
proportional transaction costs create an optimal \emph{no-trade region}: weak
signals do not justify turnover
\cite{davis1990portfolio,shreve1994optimal,delataillade2012optimal}.
A learned policy may behave similarly because it responds appropriately to
costs, lacks informative observations, or follows a poor decision rule.
Return and turnover, common endpoints in financial reinforcement learning (RL)
\cite{moody2001learning,deng2017deep,zhang2020deep,liu2021finrl},
do not by themselves distinguish these explanations.

An optimal reference offers a starting point, but its disagreements require
interpretation. A full-belief dynamic programming (DP) reference can distinguish
histories unavailable to a current-observation actor. Its action values also
assume future decisions follow the reference policy. Matching current
information leaves this second assumption unresolved: a recommended trade
may lose its advantage when subsequent decisions return to the deployed policy.

We ask: \emph{How should oracle-flagged no-trade decisions be interpreted under
deployment constraints?} Our audit examines disagreement where a frozen
proximal policy optimization (PPO) policy operates, projects reference values
onto its observations, tests isolated deviations under incumbent continuation,
and evaluates repeated deployment of observation-based alternatives. This
separates three questions: whether an oracle disagrees, whether an isolated
action change helps, and whether replacing the decision rule improves performance.

Controlled simulations show why these distinctions matter. Information matching
reduces apparent disagreement, while incumbent continuation reverses most audited
trade advantages. Repeated projected-rule deployment nevertheless improves all
50 policies. These comparisons concern different interventions; neither verdict
alone characterizes the other. Historical Bitcoin/Tether (BTCUSDT) replay
illustrates the deployment question with a hand-specified intervention, whose
relative gains arise primarily from lower turnover costs.

Our contributions are threefold. First, we provide a post-hoc audit that makes
the occupancy, information, and continuation assumptions behind no-trade
diagnosis explicit. Second, we characterize how these assumptions change the
extent and interpretation of disagreement. Third, we check diagnostic behavior
using information-representation comparisons and targeted corruptions, and
assess a frozen intervention through historical replay.

\section{Related Work}
Classical proportional-cost control derives optimal no-trade regions
\cite{davis1990portfolio,shreve1994optimal} and predictive-signal thresholds
\cite{delataillade2012optimal}. Financial RL learns trading and hedging under
frictions \cite{moody2001learning,deng2017deep,buehler2019deep,zhang2020deep};
a recent survey emphasizes noise, nonstationarity, and evaluation challenges
\cite{pippas2025evolution}. Costly-action RL learns when acting is worth its
cost \cite{mguni2023timing}, while recent partially observable trading systems
use temporal representations \cite{yu2026lstm}. We examine how to interpret
learned inaction when an optimal reference recommends a different action.

State aggregation and reactive partially observable Markov decision process
(POMDP) methods study restricted-information control
\cite{singh1994soft,li2006unified,perkins2002pomdp,
muller2022geometry,subramanian2022ais}. Recent work compresses history into
memory traces \cite{eberhard2025memory}, learns optimal reactive policies under
hard aggregation \cite{eberhard2026commit}, or uses privileged information in
training \cite{huang2025pigdreamer}. Our projection separates hidden-information
effects from within-observation disagreement using full-belief action values.
It serves diagnosis; subsequent tests evaluate specified rules without solving
for the optimal restricted-information policy.

Kalman filtering supplies the exact posterior in our linear--Gaussian setting
\cite{kalman1960new}. The lambda discrepancy detects non-Markov observations
without latent-state access \cite{allen2024lambda}; here the full belief
provides a controlled information reference. Oracle-assisted tests localize RL
bottlenecks \cite{fu2019diagnosing}. We examine the interpretation of no-trade
verdicts, using known perturbations to check diagnostic behavior and deployment
evaluation to assess candidate rules.

\section{Auditing Oracle-Flagged Inaction}
\begin{figure}[b]
\centering
\begin{tikzpicture}[
  node distance=2.8mm,
  audit/.style={draw=black!65, rounded corners=1.2pt, fill=black!3,
    text width=0.86\columnwidth, inner xsep=5pt, inner ysep=3.2pt,
    align=left, font=\normalsize},
  flow/.style={-{Latex[length=1.7mm,width=1.2mm]}, line width=0.5pt,
    draw=black!70}
]
\node[audit] (full) {\textbf{1. Full-belief audit}\\[-1pt]
  Where does the deployed policy disagree?};
\node[audit, below=of full] (proj) {\textbf{2. Current-information projection}\\[-1pt]
  Does disagreement survive information matching?};
\node[audit, below=of proj] (inc) {\textbf{3. Incumbent-continuation audit}\\[-1pt]
  Would changing this action help if PPO resumes?};
\node[audit, below=of inc] (repeat) {\textbf{4. Repeated restricted deployment}\\[-1pt]
  Does using the new rule at every step help?};
\draw[flow] (full.south) -- (proj.north);
\draw[flow] (proj.south) -- (inc.north);
\draw[flow] (inc.south) -- (repeat.north);
\end{tikzpicture}
\caption{From oracle disagreement to deployment evaluation. The first
three stages examine the reference verdict; the fourth evaluates an
observation-based replacement rule throughout deployment.}
\label{fig:audit-chain}
\end{figure}
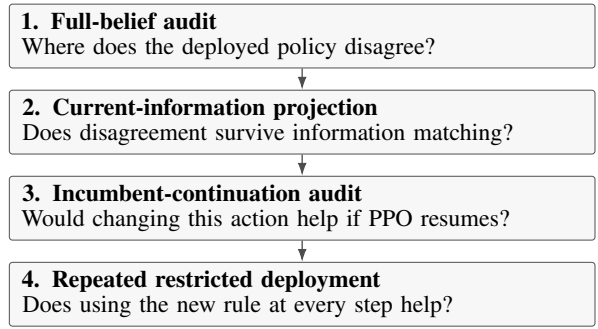

The audit asks where disagreement occurs, what information supports it,
and which future decisions make the alternative valuable.
Figure~\ref{fig:audit-chain} summarizes the comparisons for a frozen incumbent.

With feasible actions and the market fixed, we evaluate the reference on
visited states, then remove hidden-belief distinctions unavailable in the
current observation. Next, we compare a projected trade with no-trade under
incumbent continuation. Finally, we evaluate observation-measurable rules used
at every step, distinguishing policy replacement from an isolated action change.

\subsection{Controlled market and partial observation}
The latent predictive state and return obey
\begin{equation}
 z_{t+1}=\rho z_t+\xi_{t+1},\qquad
 r_{t+1}=\beta z_t+\sigma\epsilon_{t+1},
 \label{eq:dynamics}
\end{equation}
where $\xi_t\sim\mathcal N(0,\sigma_z^2)$ and
$\epsilon_t\sim\mathcal N(0,1)$ are independent. The action is the next
position $h_t\in\{-1,0,1\}$, with reward
\begin{equation}
 R_t=h_t r_{t+1}-c|h_t-h_{t-1}|-\lambda h_t^2.
 \label{eq:reward}
\end{equation}
Defaults are $\rho=0.9$, $\sigma_z=\sigma=1$, $\lambda=0$, and $\gamma=0.99$.
With $\tau=\sigma_z/\sqrt{1-\rho^2}$, we set
$S=\beta\tau/\sigma=1$ and normalized cost
$C=c/(\beta\mathbb E|z|)=1.5$ in confirmatory cohorts.

The fully observed actor receives $(z_t/\tau,h_{t-1})$. In the signal-%
processing experiment it instead observes
\begin{equation}
 x_t=z_t+\nu_t,\qquad \nu_t\sim\mathcal N(0,\sigma_\nu^2),
 \label{eq:observation}
\end{equation}
with $\sigma_\nu/\tau\in\{0.25,0.5,1\}$. The \emph{raw} actor receives
$x_t/\sqrt{\tau^2+\sigma_\nu^2}$ and no observation history. The
\emph{filtered} actor receives the standardized Kalman posterior mean. Writing
$p_t=\mathbb E[z_t\mid x_{1:t-1}]$ and using the steady-state gain $K$,
\begin{equation}
 m_t=(1-K)p_t+Kx_t,\qquad p_{t+1}=\rho m_t.
 \label{eq:kalman}
\end{equation}
In the steady-state linear--Gaussian model the posterior variance is constant,
so $m_t$ fully parameterizes the belief relevant to control.
Current-observation and filtered actors share architecture, costs, budget, and
paired seeds. A third uses the same feedforward PPO with a fixed, untuned
eight-observation standardized history.

For full observation, 101 signal bins over $[-3.5\tau,3.5\tau]$ define a finite
Markov decision process (MDP), whose optimal policy is obtained by value iteration.
The partial-%
observation reference uses the augmented Markov state $(p_t,x_t,h_{t-1})$ on
$31\times41$ and $51\times61$ prior/observation grids over $\pm3.75$ standard deviations and
the conditional reward $\beta h_tm_t-c|h_t-h_{t-1}|$. Thus all actors and the
reference face identical feasible actions, but the full-belief reference has
more information. Value iteration is
numerical within these induced finite models, not an exact continuous-state
solution. The oracle depends on $(p,x)$ through $m=(1-K)p+Kx$; retaining both
coordinates keeps the evaluation model Markov while permitting projection onto
the actor's observed cells.

\subsection{Deployment and information projection}
All diagnostics below are oracle-relative unit tests \cite{fu2019diagnosing},
comparing a frozen decision rule with a fixed feasible-action DP reference
rather than a restricted-information optimum.
Let $a^{\NT}(s)$ denote no-trade (NT), preserving the current position, and
$a^*(s)=\arg\max_aQ^*(s,a)$, with deterministic tie breaking. Define
\begin{equation}
 \Delta^*(s)=Q^*(s,a^*)-Q^*(s,a^{\NT}).
 \label{eq:oracle-gap}
\end{equation}
A state with $\Delta^*(s)>0$ is an \emph{oracle-trade opportunity}. For a
decision rule $b$, let $p_b^{\NT}(s)=b(a^{\NT}|s)$; the state is
\emph{oracle-relative false-inactive} for $b$ when both $\Delta^*(s)>0$ and
$p_b^{\NT}(s)>0.9$. Under
normalized measure $\nu$, absolute missed advantage (AMA) and weighted false
inaction (WFI) are
\begin{equation}
 \AMA_\nu(b)=\mathbb E_\nu[\Delta^*_+
 \mathbf1\{p_b^{\NT}>0.9\}],\quad
 \WFI_\nu(b)=\frac{\AMA_\nu(b)}{\mathbb E_\nu[\Delta^*_+]}.
 \label{eq:wfi}
\end{equation}
For entropy comparisons, threshold-free soft missed advantage is
$\mathbb E_\nu[\Delta^*_+p_b^{\NT}]$; its normalized form divides by
$\mathbb E_\nu[\Delta^*_+]$. One-step oracle-action regret (OAR) is
\begin{equation}
 \OAR_\nu(b)=\mathbb E_\nu[V^*(s)-\textstyle\sum_ab(a|s)Q^*(s,a)].
 \label{eq:oar}
\end{equation}
Finite-MDP linear systems give $V^b$ and stationary learned occupancy $d^b$
(the state-visitation distribution).
For full observation, the common measure is $\mu(z,h)=d_z(z)/3$; under partial
observation it is $\mu(p,x,h)=d_{\rm exo}(p,x)/3$, where $d_z$ and
$d_{\rm exo}$ are the stationary exogenous signal and joint prior--observation
distributions on the corresponding finite grids. Thus positions are uniform
and independent of the signal state; a policy is hard false-inactive when
$\WFI_\nu>0.1$ under measure $\nu$, with primary labels using $\mu$ and
deployment results $d^b$.
The probability cutoff $0.9$ and WFI cutoff $0.1$ operationalize
near-deterministic inaction and a material missed-advantage share; neither is
theoretically privileged. In the three-%
position deployment audit, all 11 labels persist over
$p_b^{\NT}\in\{0.8,0.9,0.95\}$ and WFI cutoffs $\{0.05,0.1,0.2\}$; fixed-measure
labels are probability-cutoff invariant but fall from 11 to 1 at WFI $0.2$.

For a current-observation actor, let $y=(x,h)$ denote its observed cell ($h$ is
endogenous action-state memory; no explicit $x$ history is supplied). Project
the full-belief action values using that actor's own occupancy,
$\bar Q_b(y,a)=\mathbb E_{d^b}[Q^*(s,a)\mid y]$.
\begin{equation}
 \begin{aligned}
 G_{\rm proj}^b&=\mathbb E_{d^b}[V^*-\max_a\bar Q_b(y,a)],\\
 G_{\rm cell}^b&=\mathbb E_{d^b}[\max_a\bar Q_b(y,a)
                 -\textstyle\sum_ab(a|y)\bar Q_b(y,a)],\\
 \OAR_{d^b}(b)&=G_{\rm proj}^b+G_{\rm cell}^b.
 \end{aligned}
 \label{eq:projection}
\end{equation}
The first term accounts for belief distinctions hidden within $y$; the
second measures disagreement with the action preferred by the projected values
\cite{perkins2002pomdp,muller2022geometry}. Their sum exactly decomposes one-step OAR, retaining oracle continuation
through $Q^*$. Neither term is a value gap to the optimal $y$-measurable
controller; we examine continuation next. Replacing
$Q^*$ by $\bar Q_b$ in Eqs.~\eqref{eq:oracle-gap}--\eqref{eq:wfi} gives
occupancy-projected AMA/WFI; because $d^b$ enters the conditional expectation,
each policy gets its own deployment diagnostic rather than a common restricted-%
information oracle.

\subsection{Isolated deviations and policy replacement}
To ask whether a projected trade helps when the original policy resumes,
we project the incumbent values,
$\bar Q_b^{\rm inc}(y,a)=\mathbb E_{d^b}[Q^b(s,a)\mid y]$. Let
$a_b^Y(y)$ be the best non-no-trade action under $\bar Q_b$; its matched
incumbent-continuation gap is
\begin{equation}
 \bar\Delta_b^{\rm inc}(y)=
 \bar Q_b^{\rm inc}(y,a_b^Y)-\bar Q_b^{\rm inc}(y,a^{\NT}).
 \label{eq:restricted-continuation}
\end{equation}
We report its AMA-weighted negative fraction on projected false-inactive cells.
We then deploy three deterministic candidates using only $y$. Rule
$\tilde b_*$ is greedy in $\bar Q_b$, $\tilde b_{\rm inc}$ is greedy in
$\bar Q_b^{\rm inc}$, and the common rule $\tilde b_\mu$ replaces $d^b$ by
$\mu$ in $\bar Q_b$. Exact finite-model evaluation measures each rule when repeatedly deployed.
We call this contrast the \emph{isolated--coordinated continuation gap}.
The isolated test concerns projected false-inactive cells; replacement rules
act throughout the observation space. These distinct endpoints do not establish
optimal restricted control.

The deployment endpoint is the paired reward gain
\begin{equation}
 I_J(\tilde b,b)=J(\tilde b)-J(b),
 \label{eq:deployment-gain}
\end{equation}
where both rules share the environment and evaluation convention.
Action gaps use discounted values; $J$ uses stationary net reward in simulation
and undiscounted daily net reward in BTCUSDT replay. A local advantage reversal
and $I_J>0$ therefore address distinct estimands.

\section{Experimental Protocol}
PPO uses separate two-layer 64-unit rectified linear unit actor and scalar-critic networks and
Adam ($3\!\times\!10^{-4}$). Each update collects 2,048 continuing steps; we
use four epochs, minibatches of 256, clipping 0.2, value weight 0.5, and
normalized generalized advantage estimation (GAE) advantages
\cite{schulman2016gae} with
$\lambda_{\rm GAE}=0.95$. Critic targets and rewards are not normalized.

An exploratory $9\times7\times50$ cost/entropy surface locates the fully
observed regime; a separate 50-seed anchor uses $C=1.5$, zero entropy,
and 61,440 steps. Direct continuous deployment uses 64 chains and 201/401-bin
oracle interpolation. The signal experiment uses seven arms (latent plus raw/%
filtered at three noise ratios), 50 paired seeds, and the same budget. A frozen
follow-up adds the eight-observation arm. Same-path evaluation
uses a 1,000-step burn-in and 1,000 scored steps; original arms also receive
exact stationary finite-model evaluation. Maximum residuals are
$1.71\times10^{-10}$ for Bellman equations and $1.33\times10^{-13}$ for
stationarity. Hard-incidence proportions use Wilson intervals; continuous
metrics and paired contrasts use
seed-level bootstrap intervals, averaging paths within seeds before resampling seeds.

Unit observation noise was prespecified as the sole primary raw--filtered
contrast; lower-noise cells are descriptive. Frozen actors are re-evaluated on
both partial-observation grids, and a symmetric $2\times2$ audit separates
action-rule and occupancy components. Robustness checks cover entropy $0.03$,
256,000 training steps, and a paired $S\in\{1,0.5\}$ by
$C\in\{1.5,3\}$ factorial; the long replay exactly reproduces its saved
61,440-step prefix. The restricted-continuation audit projects $Q^b$
onto the same $(x,h)$ cells as $Q^*$ and exactly evaluates three repeatedly
deployed rules with 10,000 paired bootstrap replicates. Finally, an analysis-only
check trains no policy and corrupts one endpoint component at a time: information
corruption hides prior belief while acting greedily under projected oracle values;
action corruption forces no-trade at full-belief oracle-trade states; continuation
corruption keeps the current oracle action and follows the deterministic
$Q^*$-worst rule thereafter. All use a fixed common measure on both grids.
Code and frozen policy artifacts are available from the authors upon request.

\section{Experimental Results}
We first separate action-level verdicts from deployment gains, then check
diagnostic behavior through representation comparisons and known perturbations.
Historical replay illustrates the deployment question on market data.

\subsection{From disagreement to deployment performance}
Information matching changes the extent of disagreement; continuation
changes its interpretation (Table~\ref{tab:projection}). At unit noise,
occupancy-weighted projection reduces full-belief hard labels from 49/50 to
38/50. Incumbent continuation reverses 99.3\% of projected-hard
missed-advantage mass. Yet repeated projected-rule deployment improves
all 50 policies, with mean gain 0.245 [0.205,0.284]. An unfavorable isolated-action
verdict can thus coexist with beneficial policy replacement.

\begin{table}[!ht]
\caption{Interpreting inaction at unit noise. The first three rows report
action diagnostics; the last reports the reward gain from policy replacement.}
\label{tab:projection}
\centering
\setlength{\tabcolsep}{2.2pt}
\begin{tabular}{@{}>{\raggedright\arraybackslash}p{0.40\columnwidth}
                    >{\raggedright\arraybackslash}p{0.57\columnwidth}@{}}
\toprule
Audit & Result \\
\midrule
Full belief &
49/50 hard;\newline OAR 0.293 [0.253,0.332] \\
\addlinespace[2.5pt]
Current-info projection &
38/50 hard;\newline $G_{\rm cell}=0.193$ [0.165,0.220] \\
\addlinespace[2.5pt]
Incumbent continuation &
99.3\% AMA-weighted reversal;\newline [0.989,0.996] \\
\addlinespace[2.5pt]
Repeated restricted rule &
$+0.245$ [0.205,0.284];\newline 50/50 \\
\bottomrule
\end{tabular}
\end{table}

Mean full-belief OAR 0.293 decomposes into projection and within-cell gaps
0.099 and 0.193 (residual below $10^{-14}$). The AMA-pooled reversal is 99.4\%
(without policy-level $\WFI>0.1$ conditioning); the 99.3\% in
Table~\ref{tab:projection} uses the projected-hard cohort.
A policy-independent projection gives 38/50 common-measure and 39/50 learned-%
occupancy hard labels. The incumbent-greedy rule improves all 50
policies by 0.024 [0.020,0.029]; the common-projection rule improves the mean by
0.200 [0.160,0.242] but only 39/50 policies, so it is not a per-policy dominance result.

The fully observed anchor checks whether flagged inaction persists on
states visited by the learned policy. In the high-cost, low-entropy exploratory regime corroborated
by $\OAR_\mu$, 11/50 anchor policies are hard false-inactive.
Finite-MDP deployment retains all 11, creates
none among the other 39, and gives stationary per-step regret 0.356 versus 0.030.
Direct continuous neural-actor execution preserves 11/50 labels across horizons
and post-burn-in; incumbent continuation reverses 91.9\% of missed advantage.

\subsection{Information representation and diagnostic severity}
\begin{figure}[t]
 \centering
 \includegraphics[width=\columnwidth]{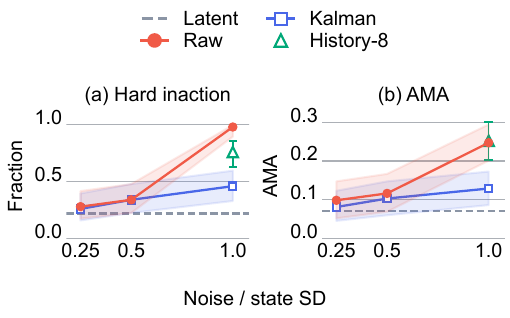}
 \caption{Learned-deployment diagnostics (50 paired seeds/arm; 95\% intervals;
 SD: standard deviation). Raw observes $x_t$; Kalman observes the posterior.
 History-8 (triangles, unit noise) lowers incidence but not AMA; Kalman reduces both.}
 \label{fig:filtering}
\end{figure}

Fewer flagged policies need not mean less missed advantage.
Table~\ref{tab:representation} compares unit-noise representations on paired
continuous paths, fixing PPO architecture, budget, costs, and seeds.
History-8 lowers hard incidence but not AMA or OAR; Kalman reduces all three.
Raw-minus-history AMA is $-0.009$ $[-0.061,0.045]$; history-minus-filtered is
0.125 [0.061,0.185]. History is limited to the fixed eight-step feedforward input;
Fig.~\ref{fig:filtering} varies noise.

\begin{table}[!ht]
\caption{Information-representation ablation at unit noise. Lower AMA and OAR
indicate less oracle-relative disagreement.}
\label{tab:representation}
\centering
\begin{tabular*}{\columnwidth}{@{\extracolsep{\fill}}lrrr@{}}
\toprule
Representation & Hard $\downarrow$ & AMA $\downarrow$ & OAR $\downarrow$ \\
\midrule
Raw & 49/50 & 0.243 & 0.293 \\
History-8 & 38/50 & 0.252 & 0.291 \\
Kalman & 23/50 & 0.127 & 0.163 \\
\bottomrule
\end{tabular*}
\end{table}

The learned-occupancy raw-minus-filtered AMA reduction, 0.117 [0.078,0.160],
comprises rule 0.064 [0.044,0.085] and occupancy 0.053 [0.033,0.073] components.
OAR components are $-0.172$ (occupancy) and 0.302 (rule).
This decomposition is descriptive, not causal.

Filtered-minus-raw gains are 0.133 [0.097,0.170] in stationary net reward and
13.43 [9.92,17.09] in common-measure discounted value, despite higher turnover
(0.076 versus 0.052; difference 0.024 [0.015,0.034]).

At entropy 0.03, filtering reduces deployment soft missed advantage by
0.126 [0.092,0.160] and raises reward by 0.144 [0.112,0.177]. At 256,000 steps,
the corresponding gains are 0.117 [0.083,0.153] and 0.130 [0.095,0.167].
Doubling normalized cost weakens the normalized soft missed-advantage benefit
under learned occupancy; signal-contrast and interaction intervals include zero.
Incumbent continuation reverses 97.2\% of residual filtered missed advantage.

\subsection{Diagnostic checks with known perturbations}
Deliberately changing information, actions, or continuation recovers all three
prespecified signatures (Table~\ref{tab:corruption-signatures}). The diagnostics
respond as intended; this does not identify training causes.
\begin{table}[!ht]
\caption{Controlled-corruption signatures on the refined grid. Recovered
entries are $G_{\rm proj}/G_{\rm cell}$ for the first two rows; continuation
reports the gap change and reversed audited advantage mass.}
\label{tab:corruption-signatures}
\centering
\setlength{\tabcolsep}{2.5pt}
\begin{tabular}{@{}p{0.20\columnwidth}p{0.43\columnwidth}p{0.27\columnwidth}@{}}
\toprule
Corruption & Expected signature & Recovered \\
\midrule
Information & $G_{\rm proj}>0,\;G_{\rm cell}=0$ & $0.180/0$ \\
\addlinespace[4pt]
Action & $G_{\rm proj}=0,\;G_{\rm cell}>0$ & $0/0.492$ \\
\addlinespace[4pt]
Continuation & Current action correct; future value reverses &
$-3.685$; $97.8\%$ \\
\bottomrule
\end{tabular}
\end{table}

The coarse grid recovers all signatures (maximum headline drift 0.0044;
numerical residual $2.77\times10^{-11}$). Refinement preserves the raw (49),
filtered (23), and projected (38) hard sets, with AMA drift below 0.003, projected-hard
reversal drift 0.0002, and candidate-gain drift below 0.001. Threshold, support,
and continuous-transition checks support these finite approximations, not the
continuous POMDP.

\subsection{A frozen intervention on BTCUSDT replay}
Historical replay illustrates Stage~4's deployment question with a hand-specified
intervention, without applying the full model-based audit.
We evaluate 50 frozen zero-entropy PPO
policies trained for 30,720 steps on
2021--2023 Binance USDT-margined BTCUSDT minute data. Training-standardized
causal inputs are taker-volume imbalance, returns, exponential moving average,
volatility, volume, and coverage. Positions $\{-1,0,1\}$ earn next-open-to-open
log returns. Daily episodes follow Coordinated Universal Time and start and
end flat; turnover costs 1.5 basis points (bp) per unit.

When PPO's hold probability exceeds 0.9 and absolute imbalance exceeds a
training-data quantile, the intervention adopts the imbalance direction if
different from inventory, holds for $H$ minutes, then resumes the hybrid rule.
Selection using 2024 data maximizes mean net gain over quantiles
$\{0.75,0.90,0.95\}$ and $H\in\{5,15,60\}$ on the prespecified 40-policy
cohort, choosing 0.75 (threshold 0.3951) and $H=60$.
Frozen for 2025, the rule applies to all 50 policies without retraining or
reselection. Replay exactly integrates action probabilities on each path.

Daily net gain is 159.11~bp in 2024 and 135.03~bp in 2025, with
47/50 and 46/50 policies improving. The 2025 common-calendar circular block
bootstrap intervals are [125.50,144.63]~bp (7 days) and [122.33,148.09]~bp
(30 days), from 2,000 replicates conditional on this fixed cohort.
In 2025, gross reward changes by $-3.71$~bp/day while costs fall by
138.74~bp/day. These cost-dominated gains over the frozen PPO cohort do not
isolate signal-based entry from the holding constraint. Always-flat earns zero
by definition; relative gains establish neither absolute profitability nor
superiority to passive holding.

\section{Conclusion}
Oracle-flagged inaction depends on information and future decisions. Our audit
distinguishes oracle disagreement, isolated action changes, and policy replacement.
In controlled experiments, information matching reduces disagreement;
incumbent continuation reverses most audited trade advantages, yet repeated
observation-based rules improve performance. Historical BTCUSDT replay
illustrates the deployment question through cost-dominated relative gains.
The audit clarifies what each comparison establishes for deployment.

\clearpage
\bibliographystyle{IEEEbib}
\bibliography{references}

@article{davis1990portfolio,
  author  = {Davis, M. H. A. and Norman, A. R.},
  title   = {Portfolio Selection with Transaction Costs},
  journal = {Mathematics of Operations Research},
  year    = {1990},
  volume  = {15},
  number  = {4},
  pages   = {676--713},
  doi     = {10.1287/moor.15.4.676}
}

@article{shreve1994optimal,
  author  = {Shreve, Steven E. and Soner, H. Mete},
  title   = {Optimal Investment and Consumption with Transaction Costs},
  journal = {The Annals of Applied Probability},
  year    = {1994},
  volume  = {4},
  number  = {3},
  pages   = {609--692},
  doi     = {10.1214/aoap/1177004966}
}

@article{delataillade2012optimal,
  author  = {de Lataillade, Joachim and Deremble, Cyril and Potters, Marc and Bouchaud, Jean-Philippe},
  title   = {Optimal Trading with Linear Costs},
  journal = {Journal of Investment Strategies},
  year    = {2012},
  volume  = {1},
  number  = {3},
  pages   = {91--115},
  doi     = {10.21314/JOIS.2012.005}
}

@inproceedings{mguni2023timing,
  author    = {Mguni, David Henry and Sootla, Aivar and Ziomek, Juliusz Krzysztof and Slumbers, Oliver and Dai, Zipeng and Shao, Kun and Wang, Jun},
  title     = {Timing Is Everything: Learning to Act Selectively with Costly Actions and Budgetary Constraints},
  booktitle = {International Conference on Learning Representations},
  year      = {2023}
}

@article{moody2001learning,
  author  = {Moody, John E. and Saffell, Matthew},
  title   = {Learning to Trade via Direct Reinforcement},
  journal = {IEEE Transactions on Neural Networks},
  year    = {2001},
  volume  = {12},
  number  = {4},
  pages   = {875--889},
  doi     = {10.1109/72.935097}
}

@article{deng2017deep,
  author  = {Deng, Yue and Bao, Feng and Kong, Youyong and Ren, Zhiquan and Dai, Qionghai},
  title   = {Deep Direct Reinforcement Learning for Financial Signal Representation and Trading},
  journal = {IEEE Transactions on Neural Networks and Learning Systems},
  year    = {2017},
  volume  = {28},
  number  = {3},
  pages   = {653--664},
  doi     = {10.1109/TNNLS.2016.2522401}
}

@article{buehler2019deep,
  author  = {Buehler, Hans and Gonon, Lukas and Teichmann, Josef and Wood, Ben},
  title   = {Deep Hedging},
  journal = {Quantitative Finance},
  year    = {2019},
  volume  = {19},
  number  = {8},
  pages   = {1271--1291},
  doi     = {10.1080/14697688.2019.1571683}
}

@article{zhang2020deep,
  author  = {Zhang, Zihao and Zohren, Stefan and Roberts, Stephen J.},
  title   = {Deep Reinforcement Learning for Trading},
  journal = {The Journal of Financial Data Science},
  year    = {2020},
  volume  = {2},
  number  = {2},
  pages   = {25--40},
  doi     = {10.3905/jfds.2020.1.030}
}

@inproceedings{liu2021finrl,
  author    = {Liu, Xiao-Yang and Yang, Hongyang and Gao, Jiechao and Wang, Christina Dan},
  title     = {{FinRL}: Deep Reinforcement Learning Framework to Automate Trading in Quantitative Finance},
  booktitle = {Proceedings of the Second ACM International Conference on AI in Finance},
  year      = {2021},
  pages     = {1--9},
  doi       = {10.1145/3490354.3494366}
}

@inproceedings{schulman2016gae,
  author    = {Schulman, John and Moritz, Philipp and Levine, Sergey and Jordan, Michael and Abbeel, Pieter},
  title     = {High-Dimensional Continuous Control Using Generalized Advantage Estimation},
  booktitle = {International Conference on Learning Representations},
  year      = {2016}
}

@inproceedings{fu2019diagnosing,
  author    = {Fu, Justin and Kumar, Aviral and Soh, Matthew and Levine, Sergey},
  title     = {Diagnosing Bottlenecks in Deep {Q}-learning Algorithms},
  booktitle = {Proceedings of the 36th International Conference on Machine Learning},
  year      = {2019},
  volume    = {97},
  series    = {Proceedings of Machine Learning Research},
  pages     = {2021--2030},
  url       = {https://proceedings.mlr.press/v97/fu19a.html}
}

@article{kalman1960new,
  author  = {Kalman, Rudolf E.},
  title   = {A New Approach to Linear Filtering and Prediction Problems},
  journal = {Journal of Basic Engineering},
  year    = {1960},
  volume  = {82},
  number  = {1},
  pages   = {35--45},
  doi     = {10.1115/1.3662552}
}

@inproceedings{muller2022geometry,
  author    = {M{\"u}ller, Johannes and Mont{\'u}far, Guido},
  title     = {The Geometry of Memoryless Stochastic Policy Optimization in Infinite-Horizon {POMDP}s},
  booktitle = {International Conference on Learning Representations},
  year      = {2022},
  url       = {https://openreview.net/forum?id=A05I5IvrdL-}
}

@inproceedings{singh1994soft,
  author    = {Singh, Satinder P. and Jaakkola, Tommi and Jordan, Michael I.},
  title     = {Reinforcement Learning with Soft State Aggregation},
  booktitle = {Advances in Neural Information Processing Systems},
  year      = {1994},
  volume    = {7},
  pages     = {361--368}
}

@inproceedings{li2006unified,
  author    = {Li, Lihong and Walsh, Thomas J. and Littman, Michael L.},
  title     = {Towards a Unified Theory of State Abstraction for {MDP}s},
  booktitle = {Proceedings of the Ninth International Symposium on Artificial Intelligence and Mathematics},
  year      = {2006},
  url       = {https://www.microsoft.com/en-us/research/wp-content/uploads/2016/02/camera-ready-9.pdf}
}

@inproceedings{perkins2002pomdp,
  author    = {Perkins, Theodore J.},
  title     = {Reinforcement Learning for {POMDP}s Based on Action Values and Stochastic Optimization},
  booktitle = {Proceedings of the Eighteenth National Conference on Artificial Intelligence},
  year      = {2002},
  pages     = {199--204},
  url       = {https://cdn.aaai.org/AAAI/2002/AAAI02-031.pdf}
}

@article{subramanian2022ais,
  author  = {Subramanian, Jayakumar and Sinha, Amit and Seraj, Raihan and Mahajan, Aditya},
  title   = {Approximate Information State for Approximate Planning and Reinforcement Learning in Partially Observed Systems},
  journal = {Journal of Machine Learning Research},
  year    = {2022},
  volume  = {23},
  number  = {12},
  pages   = {1--83},
  url     = {https://www.jmlr.org/papers/v23/20-1165.html}
}

@inproceedings{allen2024lambda,
  author    = {Allen, Cameron and Kirtland, Aaron and Tao, Ruo Yu and Lobel, Sam and Scott, Daniel and Petrocelli, Nicholas and Gottesman, Omer and Parr, Ronald and Littman, Michael L. and Konidaris, George},
  title     = {Mitigating Partial Observability in Sequential Decision Processes via the Lambda Discrepancy},
  booktitle = {Advances in Neural Information Processing Systems},
  year      = {2024},
  volume    = {37},
  doi       = {10.52202/079017-2011}
}

@inproceedings{eberhard2026commit,
  author    = {Eberhard, Onno and Vernade, Claire and Muehlebach, Michael},
  title     = {Commit to the Bit: Reactive Reinforcement Learning Done Right},
  booktitle = {Proceedings of the 43rd International Conference on Machine Learning},
  year      = {2026},
  series    = {Proceedings of Machine Learning Research},
  volume    = {306},
  url       = {https://arxiv.org/abs/2605.28276}
}

@article{pippas2025evolution,
  author  = {Pippas, Nikolaos and Turkay, Cagatay and Ludvig, Elliot Andrew},
  title   = {The Evolution of Reinforcement Learning in Quantitative Finance: A Survey},
  journal = {ACM Computing Surveys},
  year    = {2025},
  volume  = {57},
  number  = {11},
  pages   = {1--51},
  articleno = {295},
  doi     = {10.1145/3733714}
}

@article{yu2026lstm,
  author  = {Yu, Hanyue and Dong, Jiyang and Jiang, Yaqian and Sun, Xueqi},
  title   = {{LSTM}-augmented {DQN} for Quantitative Trading in Partially Observable Markets},
  journal = {Scientific Reports},
  year    = {2026},
  volume  = {16},
  articleno = {19163},
  note    = {Art.~no.~19163},
  doi     = {10.1038/s41598-026-49159-x}
}

@inproceedings{eberhard2025memory,
  author    = {Eberhard, Onno and Muehlebach, Michael and Vernade, Claire},
  title     = {Partially Observable Reinforcement Learning with Memory Traces},
  booktitle = {Proceedings of the 42nd International Conference on Machine Learning},
  year      = {2025},
  volume    = {267},
  series    = {Proceedings of Machine Learning Research},
  pages     = {14934--14949},
  url       = {https://proceedings.mlr.press/v267/eberhard25a.html}
}

@inproceedings{huang2025pigdreamer,
  author    = {Huang, Dongchi and Wang, Jiaqi and Li, Yang and Xia, Chunhe and Zhang, Tianle and Zhang, Kaige},
  title     = {{PIGDreamer}: Privileged Information Guided World Models for Safe Partially Observable Reinforcement Learning},
  booktitle = {Proceedings of the 42nd International Conference on Machine Learning},
  year      = {2025},
  volume    = {267},
  series    = {Proceedings of Machine Learning Research},
  pages     = {25859--25875},
  url       = {https://proceedings.mlr.press/v267/huang25ai.html}
}

\section{Compliance with Ethical Standards}
This study uses numerical simulations and public historical market data and
involves no human or animal subjects; no ethical approval was required.

\end{document}